\documentclass{article}

\newcommand{\bea}{\begin{eqnarray}}
\newcommand{\ena}{\end{eqnarray}}
\newcommand{\hif}{\mathchar`-}
\newcommand{\bb}{0\nu\beta\beta}
\usepackage[dvips]{color}
\usepackage{graphicx}
\begin{document}
\topmargin -1cm

\begin{flushright}
\today \\
MISC-2014-02
\end{flushright}

\begin{center}
{\large \bf 
Effects on $\sin\theta_{12}$ from perturbation of the neutrino mixing matrix with the partially degenerated neutrino masses
}\\
\vspace*{10mm}
{\sc Takeshi Araki${}^{a}$}\footnote{
E-mail: araki@cc.kyoto-su.ac.jp} and 
{\sc Eiichi Takasugi${}^{a,b}$}\footnote{
E-mail: takasugi.e@gmail.com},  

\vspace*{4mm}
${}^{a)}${\em Maskawa Institute, Kyoto Sangyo University, Kita-Ku, Kyoto 603-8555, Japan}\\
${}^{b)}${\em Department of Physics, Osaka University, Toyonaka, Osaka
 560-0043, Japan}\\
\end{center}

\begin{abstract}
We consider a situation where the leading-order neutrino mass matrix is derived by a theoretical ansatz and reproduces the experimental data well, but not completely. 
Then, the next stage is to try to fully reproduce the data by adding small perturbation terms. 
In this paper, we obtain the analytical method to 
diagonalize the perturbed mass matrix and find a consistency condition that parameters 
should satisfy not to change $\sin\theta_{12}$ much. 
This condition could cause parameter tuning and plays a crucial role in relating the added perturbation terms with the prediction analytically, in particular, for the case of the partially quasi-degenerated neutrino masses ($m_2 \simeq m_1$) where neutrinoless double beta decays would be observed in the ${\rm phase\hif I\hspace{-.1em}I}$ experiments.
\end{abstract}



\section{Introduction}
\begin{figure}[t]
\begin{center}
\includegraphics[width=12.0cm]{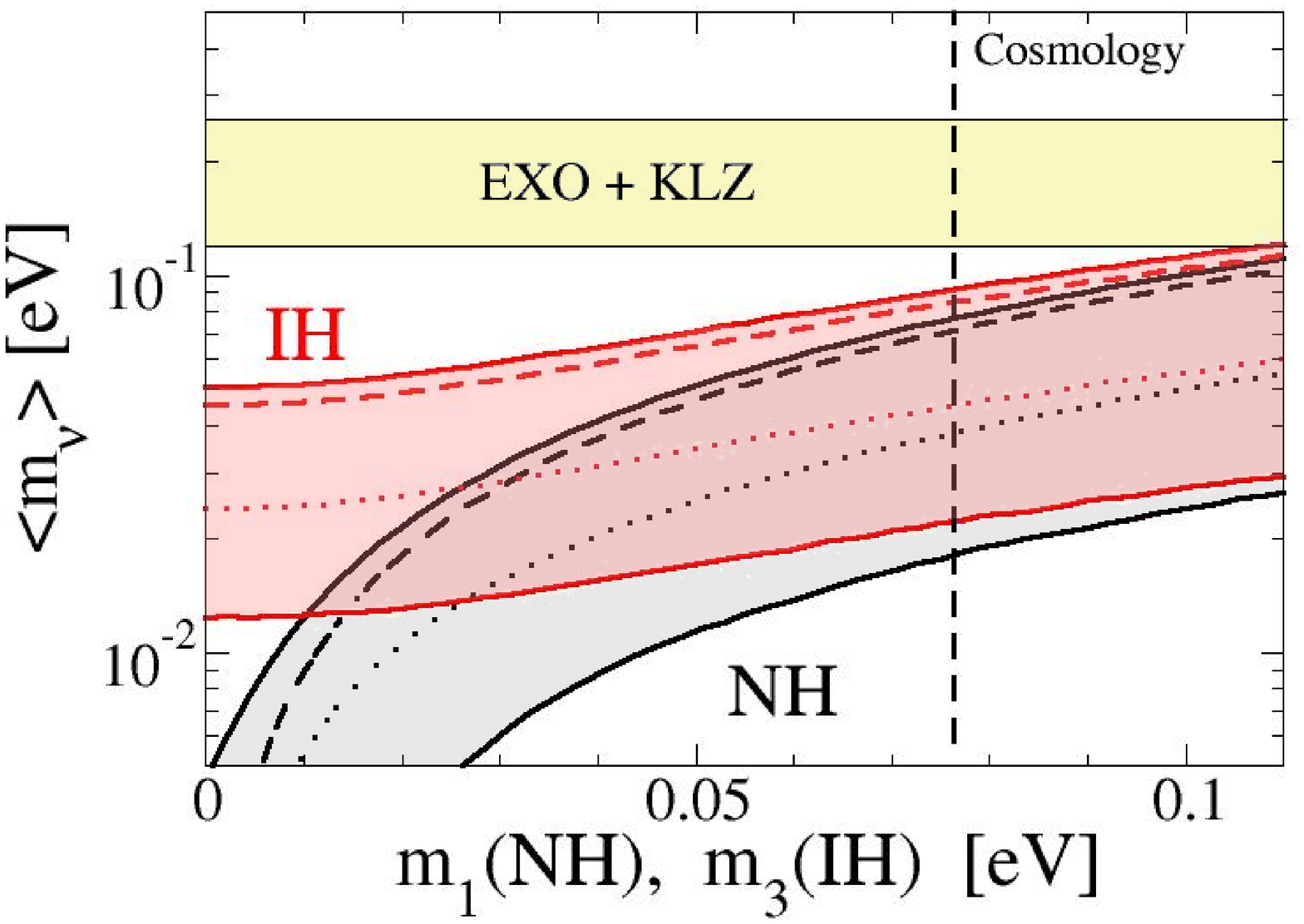}
\end{center}
\caption{\footnotesize 
The effective mass, $\langle m_\nu \rangle$, of the $\bb$ as functions of the lightest neutrino masses, $m_1$ ($m_3$) for the NH (IH) case; 
all the CP phases are varied from $0$ to $2\pi$; 
the gray (red) region is allowed by the $3\sigma$ constraints of the oscillation parameters \cite{fogli} for the NH (IH) case;
the upper (lower) regions surrounded by the dashed (dotted) and solid curves are regions of the fully constructive (destructive) interference of the Majorana phases; 
the horizontal yellow bound represents the $90\%$ C.L. upper bound on $\langle m_\nu \rangle$ from the combined analysis of the EXO and KamLAND-Zen (KLZ) experiments \cite{exo,kam}; 
the vertical dashed line corresponds to the $95\%$ C.L. upper bound on the sum of the neutrino masses from the Planck and other cosmological observations \cite{planck}.
}
\label{fig:0nbb}
\end{figure}

Various types of neutrinoless double beta decay ($\bb$) experiments have been undertaken, and 
the ${\rm phase\hif I\hspace{-.1em}I}$ experiments are planned; see Refs. \cite{0nbbR,0nbbB} for recent reviews, Refs. \cite{0nbbC1,0nbbC2,0nbbC3} for combined studies with cosmological observations, and Refs. \cite{0nbbS,0nbbP} for previous works. 
In these experiments, the expected 
sensitivity to the effective neutrino mass, $\langle m_{\nu} \rangle$, would hopefully reach $0.02~{\rm eV}$. 
As discussed by many authors, if the observed $\langle m_{\nu} \rangle$ is in regions much larger than $\sqrt{\Delta m_a^2}\simeq 0.049~{\rm eV}$, the possible mass pattern of the neutrino is the quasi-degenerate (QD) one. 
Such mass regions, however, begin to be excluded by cosmological observations (see Fig. \ref{fig:0nbb}).
If $\langle m_{\nu} \rangle$ is smaller than $0.049~{\rm eV}$, in contrast, there are several possibilities depending on the mass spectrum and the Majorana CP-violating phases \cite{bhp,sv1,sv2,doi1,doi2}.
The inverted hierarchy (IH) case with the fully constructive interference of the Majorana phases between $m_2$ and $m_1$, i.e., $\beta = \alpha$ in our notation, suggests that $\langle m_{\nu} \rangle$ is greater than or equal to $0.049~{\rm eV}$.
For the IH case with the fully destructive interference, i.e., $\beta = -\alpha$ in our notation, $\langle m_{\nu} \rangle$ is greater than or equal to $0.014~{\rm eV}$. 
In the case of the normal hierarchy (NH), with a sensitivity of $\langle m_{\nu} \rangle > 0.02~{\rm eV}$, one could explore the partially quasi-degenerated (PQD) mass regions, in which $m_2\simeq m_1$.
In particular, most of its fully constructive interference regions would be covered.
Thus, these parameter regions are expected to be important in the coming years.

As for the mixing, the observed bi-large mixing pattern \cite{pdg} motivated people to parametrize the mixing matrix with only simple numbers around the experimental data, such as Tri-Bi-Maximal (TBM) mixing \cite{TB1,TB2,TB3}.
In particular, $\sin\theta_{13}$ is predicted to be zero in these mixings.
Also, it was found that some of these mixing patterns can be derived by discrete flavor symmetries; see \cite{dfs1,dfs2,dfs3,dfs4,dfs5} for recent reviews.
Nowadays, however, these mixing patterns necessitate small perturbations because it was confirmed by long-baseline \cite{lbne1,lbne2} and reactor \cite{rct1,rct2,rct3} neutrino oscillation experiments that $\sin\theta_{13}$ is nonzero.

In this paper, with the aforementioned situation in mind, we develop an analytical method to diagonalize the perturbed neutrino mass matrix in a general way.
Let us suppose that the leading-order neutrino mass matrix 
$M_0$ is derived theoretically with using some symmetry and that its diagonalizing matrix $V_0$, which is defined by
\bea
\overline{M_0}\equiv V_{0}^T M_0 V_{0}=\pmatrix{m_1^0 e^{i\alpha_0}&0&0\cr 0&m_2^0 e^{i\beta_0}&0\cr 0&0&m_3^0\cr},
\label{eq:M0}
\ena
reproduces the experimental data of the mixing angles well, but not completely. 
Here, $m_i^0$ are taken to be real and positive, and $\beta_0$ and $\alpha_0$ are their CP phases. 
In order to fill the gap between $V_0$ and the experimental data, we add three small complex parameters:
\bea
\pmatrix{0&\epsilon_1&\epsilon_3\cr \epsilon_1&0&\epsilon_2\cr
\epsilon_3&\epsilon_2&0\cr}.
\label{eq:es}
\ena
In model-building, we put some restriction on the parameters $\epsilon_i$ and obtain a prediction.
Our question is to see analytically the relation between the restriction and the prediction. 
For this, we have to diagonalize the neutrino mass 
matrix analytically as generally as possible and then expand the exact result in terms of small parameters. 
In the course of this, we find that the parameters responsible for 
the deviations of $\sin\theta_{13}$ and $\sin\theta_{23}$ affect $\sin\theta_{12}$ as well, at the higher order of perturbation.
We also find that, in the case of the PQD mass spectrum, this effect could drastically alter $\sin\theta_{12}$.
As a result, a certain condition on the parameters is required to be satisfied in order not to change $\sin\theta_{12}$ much. 
This feature is especially prominent in the case where $(V_0)_{12}$ is very close to its experimental value.
We examine the case of TBM mixing and find that the condition causes unnatural parameter tuning for the PQD mass spectrum.
We emphasize that the condition is the result of our careful calculations.
Perturbations of a neutrino mass matrix have been widely studied by many authors in the literature \cite{pert1,pert2,pert3,pert4,pert5,pert6}.
Most of them, however, took into account only the first-order perturbation terms and/or focused on $\sin\theta_{13}$ and $\sin\theta_{23}$.
As a result, our finding has been overlooked so far.
 
In view of observability in the future $\bb$ experiments, we are mainly interested in the PQD mass regions and pay special attention to three cases: the NH with the fully constructive interference of the Majorana phases, and the IH\footnote{In the case of IH, $m_2$ is always quasi-degenerated with $m_1$.} with the fully constructive and destructive interferences.
Nevertheless, we sometimes consider the other cases for the sake of completeness.

This paper is organized as follows.
In Sect. 2, we review the behavior of $\langle m_\nu \rangle$ with respect to $p=m_2/m_3$ and the Majorana phases for the purpose of the following sections.
In Sect. 3, the diagonalization of a symmetric matrix with small perturbation terms is developed, and then the consistency condition which guarantees that $\sin\theta_{12}$ does not change much is derived in Sect. 4. 
In Sect. 5, the developed method is applied to the case of TBM mixing, and the relations between the restriction of parameters and the prediction are given for various models in Sect. 6. 
The concluding remarks are given in Sect. 7.

\section{Behavior of effective mass of $\bb$}
We use the convention that the mass parameters $m_i$ are real and positive and that $m_2$ and $m_1$ are accompanied by the Majorana phases $\beta$ and $\alpha$, respectively.
These Majorana phases appear in the mixing matrix as the phase matrix $P={\rm diag}(e^{-\frac{i}{2}\alpha},~e^{-\frac{i}{2}\beta},~1)$.
In the introduction, we argued that our main interests are 
the IH cases with both the fully constructive, $\beta=\alpha$, and destructive, $\beta=-\alpha$, interferences, and the NH case for the regions of $m_2\simeq m_1$ with the fully constructive interference. 
We summarize here the behavior of $\langle m_{\nu} \rangle$ 
for these cases. 

Let us define
\bea
p=\frac{m_2}{m_3}
\ena
\begin{itemize}
\item The NH case for the regions of $m_2\simeq m_1$ with the fully constructive interference. \\
In this case, $p<1$ and neutrino masses are expressed as
\bea
m_2=\frac{p}{\sqrt{1-p^2}}\sqrt{\Delta m_a^2},\;\;\;
m_3=\frac{1}{\sqrt{1-p^2}}\sqrt{\Delta m_a^2}.
\ena
The effective mass is written by
\bea
\langle m_{\nu} \rangle\simeq |(c_{12}c_{13})^2m_1 e^{i\alpha}+(s_{12}c_{13})^2 m_2 e^{i\beta}|
\simeq m_2=\frac{p}{\sqrt{1-p^2}}\sqrt{\Delta m_a^2},
\ena
where $s_{ij}$ ($c_{ij}$) stands for 
$\sin\theta_{ij}$ ($\cos\theta_{ij}$), and we have used $s_{13}^{} \ll 1$. 
For $\langle m_{\nu} \rangle > 0.02{\rm eV}$, one finds $p>0.4$. 

\item The IH case.\\
In this case, $p>1$ and
\bea
m_2=\frac{1}{\sqrt{1-(1/p)^2}}\sqrt{\Delta m_a^2},\;\;\;
m_3=\frac{(1/p)}{\sqrt{1-(1/p)^2}}\sqrt{\Delta m_a^2}.
\ena
On one hand, the effective mass for the fully destructive interference case is
\bea
\langle m_{\nu} \rangle\simeq m_2|c_{13}\cos 2\theta_{12}|
\simeq \frac{|\cos 2\theta_{12}|}{\sqrt{1-(1/p)^2}}\sqrt{\Delta m_a^2}
\ge 0.014 ~{\rm eV},
\ena
for the $3\sigma$ upper bound $\sin^2\theta_{12}<0.359$ \cite{fogli}.
On the other hand, the fully constructive interference case is
\bea
\langle m_{\nu} \rangle\simeq 
m_2=\frac{1}{\sqrt{1-1/p^2}}\sqrt{\Delta m_a^2}\geq 0.049~{\rm eV}.
\ena
\end{itemize}

\section{Diagonalization of symmetric matrix with small perturbation terms}
We supplement the leading-order neutrino mass matrix Eq. (\ref{eq:M0}) by the small perturbation terms in Eq. (\ref{eq:es}) and define the full mass matrix as
\bea
\overline{M}=
\mu\left[
\pmatrix{k_1&0&0\cr 0&k_2&0\cr 0&0&k_3\cr}
+
\pmatrix{0&\epsilon_1&\epsilon_3\cr \epsilon_1&0&\epsilon_2\cr
\epsilon_3&\epsilon_2&0\cr}
\right]
=\mu
\pmatrix{A&X\cr X^T&k_3},
\label{eq:Mbar}
\ena
where
\bea
A=\pmatrix{k_1&\epsilon_1\cr \epsilon_1 &k_2\cr} ,\;\;
X=\pmatrix{\epsilon_3\cr \epsilon_2\cr},
\ena 
and the overall factor $\mu$ stands for the heaviest one among $m_i^0$: $\mu= m_3^0$ ($m_2^0$) for the NH (IH) case. 
We emphasize that this is the most general complex symmetric matrix in the sense of the number of parameters.
Throughout this paper, we choose a basis in which the charged lepton mass matrix is diagonal and $k_3$ is real and positive.

We first make $\overline{M}$ block diagonalized by the unitary matrix $V_1$:
\bea
V_1=\pmatrix{u&Y^*\cr -{Y'}^T&x\cr},
\ena
where
\bea
u=\pmatrix{c_3&0\cr 
-fg^*/c_3&c_2\cr},&& x=c_3 c_2,\;\;
Y=\pmatrix{f\cr g \cr},\;\;
Y'=\pmatrix{f c_2\cr g/c_3\cr},
\ena
and 
\bea
c_3=\sqrt{1-|f|^2},\;\; c_2=\sqrt{\frac{1-|f|^2-|g|^2}{1-|f|^2}}.
\ena
This $V_1$ mainly affects $\sin\theta_{13}$ and $\sin\theta_{23}$,\footnote{
We note that there would be other unitary matrices where  $(V_{1})_{13}=f^*$ and $(V_{1})_{23}=g^*$. 
Here, we choose the one which keeps $(V_{0})_{12}$ unchanged after 
this transformation, i.e., $(V_{0}V_1)_{12} \simeq (V_{0})_{12}$.
}
and $f$ and $g$ are of the orders of $\epsilon_3$ and $\epsilon_2$, respectively, as we shall see later.
After this transformation, we find 
\bea
V_1^T\bar{M}V_1=\mu\pmatrix{K&N\cr N^T& L\cr},
\ena
where
\bea
K&=&u^T Au-u^T X{Y'}^T-Y' X^T u+k_3 Y'{Y'}^T,\nonumber\\
N&=&u^TAY^*+xu^T X-Y'X^T Y^*-k_3 xY', \nonumber\\
L&=&k_3 x^2+x(X^T Y^*+Y^{\dagger}X)+ Y^{\dagger}AY^*,
\label{eq:knl}
\ena
and $m_3=\mu|L|$.
We require that the element $N$ vanishes, which leads to
\bea
u^T X=\frac{1}{x}(k_3 xY'-u^T AY^*+Y'X^TY^*).
\label{eq:e-fg}
\ena
This identity relates the $\epsilon_i$s with $f$ and $g$, but we postpone showing their expressions until Eq. (\ref{eq:fg}).
The $2\times 2$ matrix $K$ is parametrized as
\bea
K=\pmatrix{a&c\cr c&b\cr}, 
\ena
which can be expressed explicitly in terms of the $\epsilon_i$s, 
but this is also postponed, to Eq. (\ref{eq:abc}). 
As we shall see later, $|b| \gg |c|$. 

Next, we diagonalize the matrix $K$ by the unitary matrix $V_2$:
\bea
V_2=
\pmatrix{{\cal C}& {\cal S} e^{i\kappa}\cr
 -{\cal S} e^{-i\kappa}&{\cal C} \cr},
\ena
where ${\cal C}=\cos\Theta$ and ${\cal S}=\sin\Theta$, and $V_2$ affects $\theta_{12}$. 
The important point is that ${\cal S}$ will be much smaller than  $f$ and $g$, 
because we assume that $(V_0)_{12}$ is very close to the experimental data.
The angle $\Theta$ and the phase $\kappa$ are given by 
\bea
\tan 2\Theta=2\frac{|a^*c+bc^*|}{|b|^2-|a|^2},\;\;\;
\kappa={\rm arg}(a^*c+bc^*),
\label{eq:tan2}
\ena
respectively.
The eigenvalues are found to be
\bea
&&|\lambda_1|^2 = \left( \frac{m_1}{\mu} \right)^2
= \frac{1}{2}\left\{ |a|^2 + |b|^2 + 2|c|^2 - \frac{|b|^2 - |a|^2}{\cos 2\Theta}\right\}
, \nonumber\\
&&|\lambda_2|^2 = \left( \frac{m_2}{\mu} \right)^2
= \frac{1}{2}\left\{ |a|^2 + |b|^2 + 2|c|^2 +\frac{|b|^2 - |a|^2}{\cos 2\Theta}\right\} ,
\label{eq:lam}
\ena
where $\mu$ is the overall factor defined in Eq. (\ref{eq:Mbar}), and $m_{1,2}$ are the physical neutrino masses, which are real and positive.
From them, the mass splitting between $m_2$ and $m_1$ is written as 
\bea
|\lambda_2|^2 -|\lambda_1|^2
=\frac{\Delta m_s^2}{\mu^2}
=\frac{|b|^2 - |a|^2}{\cos 2\Theta},
\ena
and we find
\bea
\sin 2\Theta
=2\mu^2 \frac{|a^*c+bc^*|}{\Delta m_s^2}.
\label{eq:sin2}
\ena
The neutrino mixing matrix is obtained by $(V_0V_1V_2)$ aside from 
phases of neutrino masses, which are related to the Majorana phases.

Up to now, the analysis is exact. 
In what follows, we exploit the fact that $\epsilon_3$ and $\epsilon_2$ (thus, $f$ and $g$) are small and that $\epsilon_1$ is much smaller than them: as we shall show later, $\epsilon_1$ should be of the order of $\epsilon_{2,3}^2$ or much smaller than it.
We hereafter omit terms which are higher than $f^2$, $g^2$ and terms proportional to $\epsilon_2 f$ and $\epsilon_2 g$. 
In this case, Eq.(\ref{eq:e-fg}) reduces to be
\bea
X \simeq k_3 Y-A Y^*,
\label{eq:e-fg2}
\ena
yielding
\bea
f \simeq \frac{1}{k_3^2 - |k_1|^2}
\left[
k_3^{} \epsilon_3 + k_1\epsilon_3^*
\right],~~~~
g \simeq \frac{1}{k_3^2 - |k_2|^2}
\left[
k_3^{} \epsilon_2 + k_2\epsilon_2^*
\right],
\label{eq:fg}
\ena
or
\bea
\epsilon_3  \simeq k_3 f-k_1 f^*, ~~~~
\epsilon_2  \simeq k_3 g-k_2 g^*,
\ena
and the parameters $a$, $b$, and $c$ included in $K$ are expressed as
\bea
a&\simeq& k_1(1+|f|^2)-k_3f^2, \nonumber\\
b&\simeq& k_2(1+|g|^2)-k_3g^2, \nonumber\\
c&\simeq& \epsilon_1 -\epsilon_3 g .
\label{eq:abc}
\ena 
Now, we can compute in a good approximation the neutrino mixing matrix 
$V=(V_0V_1V_2)$, once $V_0$ is given. \\

\section{Consistency conditions}
One may think that the mixing angles are only moderately corrected since the $\epsilon$s are assumed to be small.
However, ${\cal S}$ is not necessarily small; rather, it could take an unrealistically large value.
This is because the denominator of Eq. (\ref{eq:sin2}) is precisely measured and is very small.
In order for the full mixing matrix $V$ to be consistent with the experimental data, therefore, one needs to somehow make the numerator sufficiently small, which leads to
\bea
|a^*c+bc^*| \simeq \frac{\Delta m_s^2}{\mu^2}~ \Theta \simeq 0.
\label{eq:cc}
\ena
We hereafter refer to this requirement as the {\it consistency condition}.
In the following, we further examine it by categorizing the neutrino mass spectrum into three types.

\begin{enumerate}
\item The NH case in the regions of $m_2 \gg m_1$.\\
In the case of NH, $\mu=m_3^0$ and
\bea
k_1=\frac{m_1^0}{m_3^0}e^{i\alpha_0},\;\;\;
k_2=\frac{m_2^0}{m_3^0}e^{i\beta_0},\;\;\; 
k_3=1.
\ena
With Eq. (\ref{eq:abc}), the left-hand side of Eq. (\ref{eq:cc}) is written by 
\bea
|a^*c + bc^*| \simeq
\left|
\frac{m_1^0}{m_3^0}e^{-i\alpha_0}(\epsilon_1-\epsilon_3 g)
+\frac{m_2^0}{m_3^0}e^{i\beta_0}(\epsilon_1-\epsilon_3 g)^*
\right|.
\ena
Note that $m_i^0$ are taken to be real and positive, and $g$ is given in Eq. (\ref{eq:fg}).
Since $m_2 \simeq m_2^0$ and $m_2 \gg m_1$, the term proportional to $m_1^0$ may be dropped in comparison with that of $m_2^0$.
By using the approximations $m_3^0 \simeq m_3 \simeq \sqrt{\Delta m_a^2}$ and $p_0 = m_2^0/m_3^0 \simeq p\simeq \sqrt{\Delta m_s^2/\Delta m_a^2}$, we find 
\bea
\left|\epsilon_1-\epsilon_3 g\right|
\simeq \sqrt{\frac{\Delta m_s^2}{\Delta m_a^2}}~\Theta \simeq 0.
\label{eq:ccnh1}
\ena

\item The NH case in the regions of $m_2 \simeq m_1$ ($p > 0.4$).\\ 
This case occurs when the neutrinoless double beta decay is observed in the ${\rm phase\hif I\hspace{-.1em}I}$ experiments.
By taking the limit of $m_i = m_i^0$ and $m_2^0 = m_1^0$, the consistency condition can be rewritten as 
\bea
\left|{\rm Re}(e^{-\frac{i}{2}(\alpha_0+\beta_0)}
(\epsilon_1-\epsilon_3 g))\right|
\simeq \frac{(1-p^2)}{2p}\left(\frac{\Delta m_s^2}{\Delta m_a^2} \right)~\Theta \simeq 0.
\label{eq:ccnh2}
\ena
Note that $\beta_0 \simeq \beta$ and $\alpha_0 \simeq \alpha$.

\item The IH case.\\
In the case of IH, $\mu=m_2^0$ and 
\bea
k_1=\frac{m_1^0}{m_2^0}e^{i\alpha_0},\;\;\;
k_2=e^{i\beta_0},\;\;\;
k_3=\frac{m_3^0}{m_2^0}.
\ena
Since $m_2$ is always quasi-degenerated with $m_1$, the consistency condition turns out to be
\bea
\left|{\rm Re}\left(e^{-\frac{i}{2}(\alpha_0+\beta_0)}
(\epsilon_1-\epsilon_3 g)\right)\right|
\simeq \frac{(1-(1/p)^2)}{2}\left(\frac{\Delta m_s^2}{\Delta m_a^2} \right)~\Theta \simeq 0.
\label{eq:ccih}
\ena
\end{enumerate}
In all the cases, the key ingredient is $\epsilon_1 - \epsilon_3 g$, and the consistency conditions force $\epsilon_1$ to be of the order of $\epsilon_{2,3}^2$.
In other words, one needs to tune $\epsilon_1$ to cancel out $\epsilon_3 g$.
As we shall demonstrate in the next section, this causes unnatural parameter tuning in some cases.

\section{Tri-bi-maximal mixing case}
We here choose the TBM mixing matrix $V_{\rm TBM}$ as $V_0$,
\bea
V_{\rm TBM}=\pmatrix{\sqrt{\frac{2}{3}}&\frac{1}{\sqrt{3}}&0\cr 
-\frac{1}{\sqrt{6}}&\frac{1}{\sqrt{3}}&\frac{1}{\sqrt{2}}\cr 
\frac{1}{\sqrt{6}}&-\frac{1}{\sqrt{3}}&\frac{1}{\sqrt{2}}\cr}.
\ena
In this case, the full mixing matrix after perturbation is obtained as
\bea
V&=&V_{\rm TBM}V_1 V_2 \nonumber\\
&\simeq &\pmatrix{
\sqrt{\frac{2}{3}}\left(1-\frac{1}{\sqrt{2}}{\cal S} e^{-i\kappa}\right) & 
\frac{1}{\sqrt{3}}\left(1+\sqrt{2}{\cal S} e^{i\kappa}\right)
& \frac{1}{\sqrt{3}}(\sqrt{2}f+g)^*\cr
-\frac{1}{\sqrt{6}}(1+\sqrt{3}f- \sqrt{2}{\cal S} e^{-i\kappa})&
\frac{1}{\sqrt{3}}\left(1-\sqrt{\frac{3}{2}}g-\frac{1}{\sqrt{2}}{\cal S} e^{i\kappa} \right)
&\frac{1}{\sqrt{2}}\left(1+\frac{1}{\sqrt{3}}(-f+\sqrt{2}g)^* \right)\cr
\frac{1}{\sqrt{6}}(1-\sqrt{3}f+\sqrt{2}{\cal S} e^{-i\kappa})&
-\frac{1}{\sqrt{3}}\left(c_2+\sqrt{\frac{3}{2}}g-\frac{1}{\sqrt{2}}{\cal S} e^{i\kappa}\right)&
\frac1{\sqrt{2}}\left(1-\frac{1}{\sqrt{3}}(-f+\sqrt{2}g)^*\right)\cr},
\nonumber
\ena 
up to the first order of $f$, $g$, and ${\cal S}$.
The mixing angles are derived as
\bea
&&\sin\theta_{13}~e^{-i\delta} \simeq V_{13}
= \frac{1}{\sqrt{3}} (\sqrt{2}f+g )^*, 
\nonumber \\
&&\sin^2\theta_{23}
\simeq |V_{23}|^2
\simeq \frac{1}{2}\left( 1+\frac{2}{\sqrt{3}}{\rm Re}[-f+\sqrt{2}g] \right),
\label{eq:tbm-angles}
\ena
and
\bea
\sin^2\theta_{12}
= \frac{|V_{12}|^2}{c_{13}^2}
\simeq 
\frac{1}{3}
\left(
1 + 2\sqrt{2}{\cal S}\cos\kappa + {\cal S}^2 - 
\frac{2}{3}\left\{ |g|^2 - |f|^2 - \sqrt{2}{\rm Re}[fg^*] \right\}
\right),
\label{eq:tbm-s12}
\ena 
where $\kappa$ is defined in Eq. (\ref{eq:tan2}).
We have taken into account the second-order terms of $f$ and $g$ for $\sin^2\theta_{12}$ as they could be the first correction terms depending on the sizes of $\cos\kappa$ and ${\cal S}$.
Note that the orders of $|f|$ and $|g|$ are constrained by $\sin\theta_{13}$ and $\sin\theta_{23}$, and their contributions via the fourth term to $\sin^2\theta_{12}$ are at most $\pm 0.01$; in contrast, they are crucial when evaluating ${\cal S}$, as we outlined in Sect. 3.

According to the latest global analysis by Capozzi et al. \cite{fogli}, the allowed $2\sigma$ ($3\sigma$) range is $0.275(0.259)\le \sin^2 \theta_{12} \le 0.342(0.359)$, which places
\bea
F-0.062(-0.079) ~~\le~~ {\cal S}\left[ \cos\kappa + \frac{1}{2\sqrt{2}}{\cal S}\right] ~~\le~~ F+0.009(0.027) ,
\ena
where $F=\sqrt{2}/6\left\{ |g|^2 - |f|^2 - \sqrt{2}{\rm Re}[fg^*]\right\}$.
The angle ${\cal S}\simeq \Theta$ is much smaller than the first-order term as long as $\cos \kappa$ is not very small. 
Even in the case $\cos \kappa=0$, ${\cal S}$ is the first-order term.  

Below, we examine the behavior of $\cos \kappa$ for the three cases defined in Sect. 4.
\begin{enumerate}
\item The NH case in the regions of $m_2 \gg m_1$.\\
We find $a^*c+bc^*\simeq p_0 e^{i\beta_0}(\epsilon_1-\epsilon_3 g)^*$, so that
\bea
\kappa \simeq \beta_0-{\rm arg}(\epsilon_1 - \epsilon_3 g).
\ena
It may be worthwhile to note that $\beta_0$ is almost equal to the Majorana CP-violating phase $\beta$ because the phase of $V_{12}$ is suppressed and phases of $V_{23}$ and $V_{33}$ are absorbed by charged lepton fields.  

\item The NH case in the regions of $m_2 \simeq m_1$.\\
Since $m_1^0 \simeq m_2^0$, we find 
\bea
a^*c+bc^* \simeq 2p_0 e^{-\frac{i}{2}(\alpha_0-\beta_0)}
{\rm Re}\left[e^{-\frac{i}{2}(\alpha_0+\beta_0)}(\epsilon_1 - \epsilon_3 g)\right],
\ena
and thus
\bea
\kappa \simeq -\frac{1}{2}(\alpha_0 - \beta_0).
\ena
Because $\alpha_0 \simeq \alpha$ and $\beta_0 \simeq \beta$, one readily notices that
\bea
&&\cos\kappa \simeq \pm 1 ~~~{\rm when}~~~\alpha \simeq \beta~~{\rm or}~~\alpha \simeq \beta+2\pi, \nonumber\\
&&\cos\kappa \simeq 0 ~~~{\rm when}~~~\alpha \simeq \beta \pm \pi.
\ena
Namely, the former happens in the case of the fully constructing interference of the Majorana phases, while the latter is the case of the fully destructive interference.
It should be noted that for $\cos\kappa=-1$, the correction decreases $\sin\theta_{12}$ because we choose ${\cal S} \ge 0$, while for $\cos \kappa=1$ and $\cos \kappa=0$, the correction increases it. 
The present tendency seems to disfavor the $\cos\kappa=1$ and $\cos \kappa=0$ cases.

\item The IH case.\\ 
In this case, we find
\bea
a^*c+bc^*\simeq 2 e^{-\frac{i}{2}(\alpha_0-\beta_0)}
{\rm Re}\left[ e^{-\frac{i}{2}(\alpha_0+\beta_0)}(\epsilon_1 - \epsilon_3 g) \right],
\ena
and thus
\bea
\kappa \simeq -\frac{1}{2}(\alpha_0 - \beta_0).
\ena
Note that $\alpha_0$ and $\beta_0$ are not necessarily equal to the physical Majorana phases when $m_3 \simeq 0$, but $\alpha_0 - \beta_0 \simeq \alpha - \beta$ still holds.\footnote{
Also, if $\alpha_0 \simeq \beta_0$, then $\alpha \simeq \beta$.
}
Therefore, like the previous case, $\cos\kappa \simeq \pm 1$ and $\cos\kappa \simeq 0$ occur in the fully constructive and destructive interference cases, respectively. 
\end{enumerate}

\subsection{Parameter tuning}
Let us roughly estimate how strong the parameter tuning required by the consistency condition is.
Taking the limits of $\cos\kappa= -1$ and $\cos\kappa = 0$, we place $|\sin^2\theta_{12}^{}-1/3| \le 0.025$.
This number corresponds to the best-fit value and $3\sigma$ upper bound \cite{fogli} for $\cos\kappa= -1$ and $0$, giving rise to $\Theta \le 0.027$ and $0.28$, respectively. 
Also, we will use $\Delta m_s^2 / \Delta m_a^2 = 0.031$ and ignore the fourth term in Eq. (\ref{eq:tbm-s12}).
\begin{enumerate}
\item The NH case in the regions of $m_2 \gg m_1$.\\
The consistency condition is given in Eq. (\ref{eq:ccnh1}).
For $\cos\kappa = -1$, we find
\bea
\left| \frac{\epsilon_1}{\epsilon_3 g} -1 \right|
< 0.12.
\ena
For $\cos\kappa = 0$, the parameter tuning is not so serious.
We have substituted $|\epsilon_3 g|\simeq |fg| =0.04$ in view of $\sin^2\theta_{13}^{\rm best} \simeq 0.023$ \cite{fogli}.

\item The NH case in the regions of $m_2 \simeq m_1$.\\
We simplify the left-hand side of Eq. (\ref{eq:ccnh2}) as $|\epsilon_1-\epsilon_3 g|$.
For $\cos\kappa = -1$ and $\cos\kappa = 0$, we find
\bea
\left| \frac{\epsilon_1}{\epsilon_3 g} -1 \right|
< 0.037 (0.023)~~{\rm for}~~p=0.4 (0.8),
\ena
and
\bea
\left| \frac{\epsilon_1}{\epsilon_3 g} -1 \right|
< 0.37 (0.24)~~{\rm for}~~p=0.4 (0.8),
\ena
respectively, where $p=m_2/m_3$, and $|\epsilon_3 g|\simeq (1-p)|fg| =0.04(1-p)$ is assumed.

\item The IH case.\\
We simplify the left-hand side of Eq. (\ref{eq:ccih}) as $|\epsilon_1-\epsilon_3 g|$.
For $\cos\kappa = -1$ and $\cos\kappa = 0$, we find
\bea
\left| \frac{\epsilon_1}{\epsilon_3 g} -1 \right|
< 0.010 (0.019)~~{\rm for}~~1/p=0 (0.8),
\ena
and
\bea
\left| \frac{\epsilon_1}{\epsilon_3 g} -1 \right|
< 0.11 (0.19)~~{\rm for}~~1/p=0 (0.8),
\ena
respectively, where $|\epsilon_3 g|\simeq (1-1/p)|fg| =0.04(1-1/p)$ is assumed.
\end{enumerate}
As demonstrated above, from a few $\%$ to several tens of $\%$ tuning is required between $\epsilon_1$ and $\epsilon_3 g$.
In particular, somewhat strong parameter tuning may be necessary in the case of $m_2 \simeq m_1$ with $\beta \simeq \alpha$.

\subsection{Validity of consistency conditions}
\begin{figure}[t]
\begin{center}
\includegraphics[width=7.0cm]{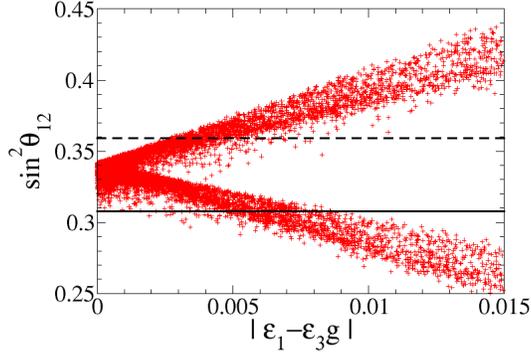}
\end{center}
\caption{\footnotesize 
Scatter plot of $\sin^2\theta_{12}$ for the NH in the case of $m_1 = 0$ and $\cos\kappa = \pm 1$.
The horizontal dashed and solid lines display the $3\sigma$ upper bound and best-fit value, respectively.
}
\label{fig:ccnh1}
\end{figure}

\begin{figure}[t]
\begin{center}
\includegraphics[width=7.0cm]{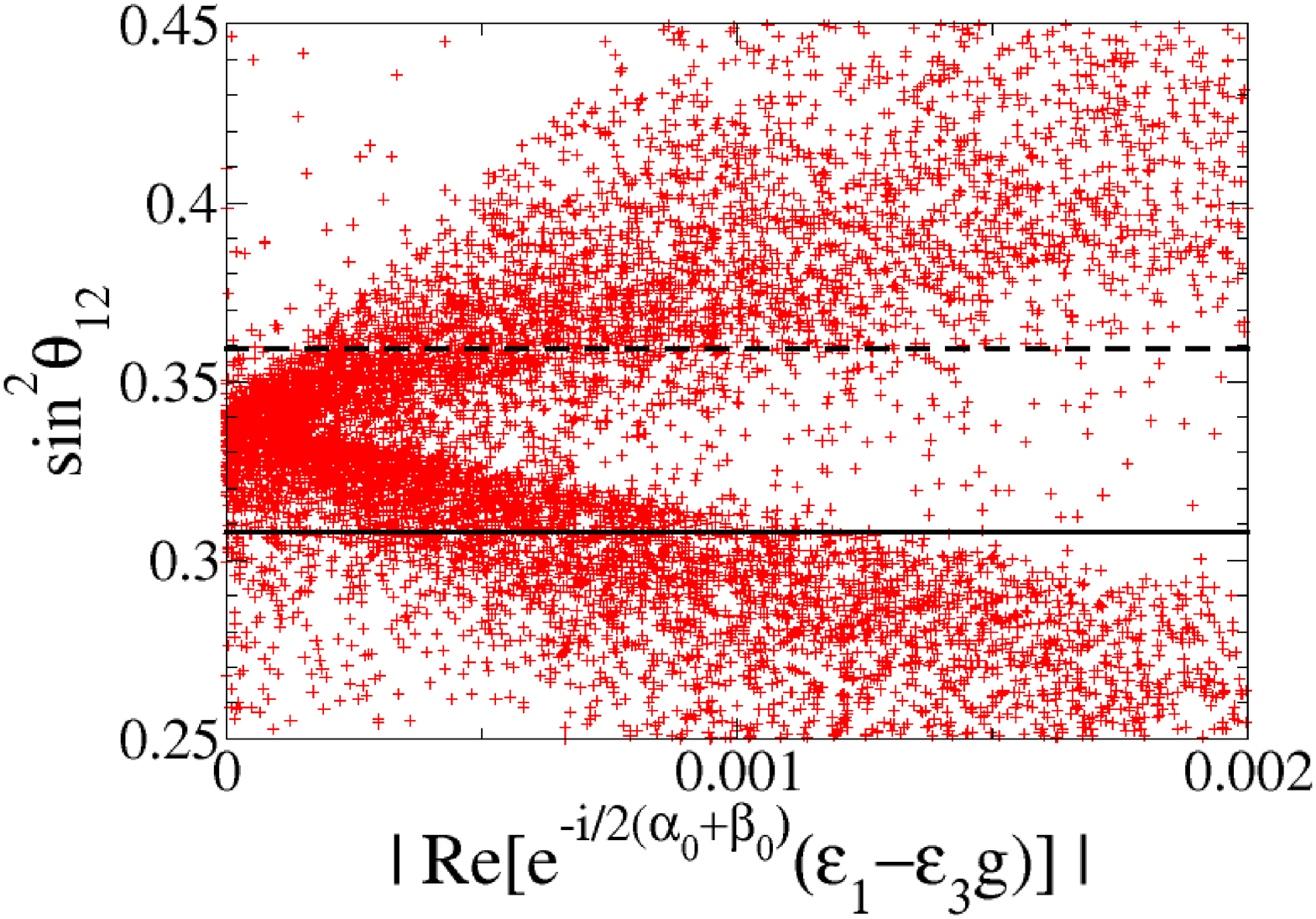}
\hspace{1.0cm}
\includegraphics[width=7.0cm]{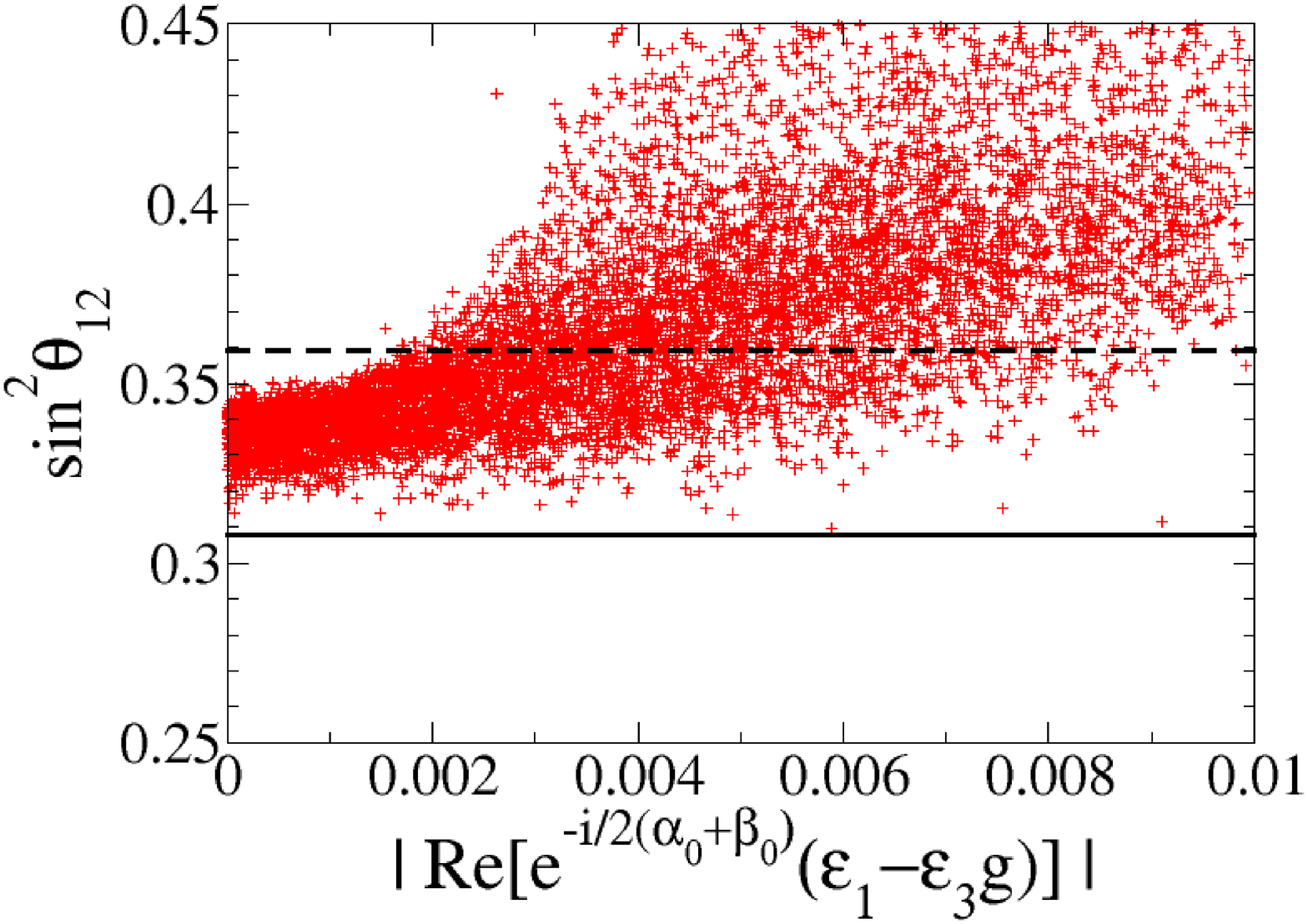}
\end{center}
\caption{\footnotesize 
Scatter plots of $\sin^2\theta_{12}$ for the NH in the mass regions of $m_2 \simeq m_1$ ($p=0.4 - 0.8$), with $\cos\kappa = \pm 1$ (left panel) and $\cos\kappa =0$ (right panel).
The horizontal dashed and solid lines display the $3\sigma$ upper bound and best-fit value, respectively.
}
\label{fig:ccnh2}
\end{figure}

We numerically diagonalize the mass matrix and check the validity of the consistency conditions Eqs. (\ref{eq:ccnh1}), (\ref{eq:ccnh2}), and (\ref{eq:ccih}).
In the numerical calculations, we place the $1\sigma$ error bounds for $\Delta m_s^2$, $\Delta m_a^2$, $\sin^2\theta_{13}$, and $\sin^2\theta_{23}$ from Ref. \cite{fogli}:
\bea
&&\Delta m_s^2=(7.32 - 7.80)\times 10^{-5}~~{\rm eV}^2,~~
\Delta m_a^2=
\left\{\begin{array}{l}
(2.38 - 2.52)\times 10^{-3}~~{\rm eV}^2 \\
(2.33 - 2.47)\times 10^{-3}~~{\rm eV}^2
\end{array}\right. , \nonumber \\
&&\sin^2\theta_{13}=
\left\{\begin{array}{l}
(2.16 - 2.56)\times 10^{-2} \\
(2.18 - 2.60)\times 10^{-2}
\end{array}\right. ,
~~\sin^2\theta_{23}=
\left\{\begin{array}{l}
(3.98 - 4.54)\times 10^{-1}~~~{\rm for}~~{\rm NH} \\
(4.08 - 4.96)\times 10^{-1}~~~{\rm for}~~{\rm NH}
\end{array}\right. .
\ena
In Figs. \ref{fig:ccnh1} and \ref{fig:ccnh2}, we plot $\sin^2\theta_{12}$ as a function of the left-hand side of the consistency condition for Eqs. (\ref{eq:ccnh1}) and (\ref{eq:ccnh2}).
The figures for Eq. (\ref{eq:ccih}) are almost the same as Fig. \ref{fig:ccnh2}.
In Fig. \ref{fig:ccnh1}, $m_1 = 0$ and $\cos\kappa = \pm 1$ are assumed.
The left and right panels in Fig. \ref{fig:ccnh2} are the cases of the fully constructive interference ($\cos\kappa = \pm 1$) and destructive interference ($\cos\kappa = 0$), respectively, for $p=0.4 - 0.8$.
All the CP phases are varied from $0$ to $2\pi$, and $|\epsilon_2|$ and $|\epsilon_3|$ run from $0.00$ to $0.25$. 

From the figures, one can observe a trend that $\sin^2\theta_{12}$ approaches its TBM value as the consistency conditions are satisfied.
In the left panel of Fig. \ref{fig:ccnh2}, however, $\sin^2\theta_{12}$ departs from the TBM value even if the $x$-axis is zero.
This is due to the failure of the approximations made above Eq. (\ref{eq:e-fg2}), and this indicates that one needs to take into account the next higher-order terms and tune $\epsilon_1 - \epsilon_3 g$ to cancel them out.
The resulting condition would be very complex and require much more delicate parameter tuning.
Hence, we do not go into its detail here.
In the next section, we shall invent several models where the consistency conditions Eqs. (\ref{eq:ccnh2}) and (\ref{eq:ccih}) work very well.

\section{Applications to models}
As we demonstrated in the previous section, the consistency conditions could be satisfied by tuning $\epsilon_1$.
However, it may be difficult to explain such parameter tuning by model-building.
Furthermore, in some cases, the consistency conditions fail to keep $\sin\theta_{12}$ within experimentally realistic ranges.
In this section, we consider two other possibilities by postulating $\epsilon_1 = 0$: (1) adjusting either $|\epsilon_2|$ or $|\epsilon_3|$ to be very small, and (2) adjusting CP phases.
In the models proposed below, the consistency conditions work very well.
Moreover, they seem attractive from model-building and/or phenomenological points of view.
For definition, we again employ TBM mixing as $V_0$.

\subsection{Adjusting $|\epsilon_2|$ or $|\epsilon_3|$}
The consistency conditions Eqs. (\ref{eq:ccnh1}), (\ref{eq:ccnh2}), and (\ref{eq:ccih}) can be satisfied by making either $|\epsilon_2|$ or $|\epsilon_3|$ vanish when $|\epsilon_1| =0$.
The following arguments are independent of the neutrino masses and Majorana phases. 
\begin{itemize}
\item $|\epsilon_2|=0$ ($g=0$) case. \\
In this case, the mixing matrix turns out to be the so-called tri-maximal mixing \cite{triM}:
\bea
V=V_{\rm TBM}V_1(\epsilon_2=0)
\simeq \pmatrix{
\sqrt{\frac{2}{3}} & 
\frac{1}{\sqrt{3}}
& \sqrt{\frac{2}{3}}f^*\cr
-\frac{1}{\sqrt{6}}(1+\sqrt{3}f)&
\frac{1}{\sqrt{3}}
&\frac{1}{\sqrt{2}}\left(1-\frac{1}{\sqrt{3}}f^*\right)\cr
\frac{1}{\sqrt{6}}(1-\sqrt{3}f)&
-\frac{1}{\sqrt{3}}&
\frac1{\sqrt{2}}\left(1+\frac{1}{\sqrt{3}}f^*\right)\cr}.
\ena
Its mixing properties have been extensively studied by many authors, so we refrain from going into details.
See, for instance, Refs. \cite{modTB1,stw,modTB2} for the behavior of $\sin^2\theta_{12}$ and the others.
Nevertheless, several comments are in order.
(1) The higher-order term of $f$ included in Eq. (\ref{eq:tbm-s12}) slightly increases $\sin^2\theta_{12}$; thus $\sin^2\theta_{12} > 1/3$ is predicted.
(2) The model has a prediction involving $\sin\theta_{13}$, $\sin\theta_{23}$, and the Dirac phase $\delta$:
\bea
\cos\delta=\frac{\sqrt{2}}{\sin \theta_{13}}
\left(\frac{1}{2}-\sin^2\theta_{23} \right).
\label{eq:efm}
\ena
Note that when $\sin\theta_{23}=1/\sqrt{2}$, then $\delta=\pi/2$ or $3\pi/2$.

It may be interesting to note that the mass matrix preserves a $Z_2$ symmetry even after adding the perturbation terms.
It is well known that TBM mixing can be derived from the mass matrix invariant under the following $Z_2$ symmetries \cite{Lam1,Lam2,Lam3} (see also Ref. \cite{GLL}):
\bea
G_1^{\rm TBM}=
\pmatrix{
 1 & 0 & 0 \cr
 0 & 0 &-1 \cr
 0 &-1 & 0 \cr
},~~~
G_2^{\rm TBM}=\frac{1}{3}
\pmatrix{
 1 &-2 & 2 \cr
-2 & 1 & 2 \cr
 2 & 2 & 1 \cr
},
\ena
in the flavor basis.
In the case of $|\epsilon_1|=|\epsilon_2|=0$, $G_2^{\rm TBM}$ remains unbroken.
This often happens in a class of the $A_4$ flavor model \cite{AF1,AF2,Ma} because $A_4$ does not include $G_1^{\rm TBM}$.

It should also be noted that the difficulty in keeping $\theta_{12}$ around the TBM value while reproducing a large $\theta_{13}$ in the $A_4$ flavor model was pointed out in Refs. \cite{Lin,AF3,AF4}.
They arrived at the same solution, $\epsilon_1 = \epsilon_2 = 0$, and Eq. (\ref{eq:efm}).

\item $|\epsilon_3|=0$ ($f=0$) case.\\ 
In this case, the mixing matrix takes the form of
\bea
V=V_{\rm TBM}V_1(\epsilon_3=0)
\simeq \pmatrix{
\sqrt{\frac{2}{3}} & 
\frac{1}{\sqrt{3}}&
 \frac{1}{\sqrt{3}}g^*\cr
-\frac{1}{\sqrt{6}}&
\frac{1}{\sqrt{3}}\left(1-\sqrt{\frac{3}{2}}g \right)
&\frac{1}{\sqrt{2}}\left(1+\sqrt{\frac{2}{3}}g^* \right)\cr
\frac{1}{\sqrt{6}}&
-\frac{1}{\sqrt{3}}\left(1+\sqrt{\frac{3}{2}}g \right)&
\frac1{\sqrt{2}}\left(1-\sqrt{\frac{2}{3}}g^* \right)\cr
}.
\ena 
This mixing patten is also analyzed in Refs. \cite{modTB1,modTB2}.
In contrast to the tri-maximal mixing, the higher-order term of $g$ included in Eq. (\ref{eq:tbm-s12}) slightly decreases $\sin^2\theta_{12}$; thus $\sin^2\theta_{12} < 1/3$ is predicted.
The model prediction among $\sin\theta_{13}$, $\sin\theta_{23}$, and $\delta$ is 
\bea
\cos\delta=\frac{1}{\sqrt{2}\sin\theta_{13}} \left( \sin^2 \theta_{23}-\frac{1}{2} \right).
\ena
As in the case of the tri-maximal mixing, the mass matrix preserves $G_1^{\rm TBM}G_2^{\rm TBM}$ in the flavor basis.
\end{itemize}

\subsection{Adjusting phases}
We restrict ourselves to the case of $m_2 \simeq m_1$ (as well as $\epsilon_1 = 0$) and parametrize $\epsilon_3$ and $\epsilon_2$ as 
\bea
\epsilon_3=E_3e^{i\rho_3},\;\;\; \epsilon_2=E_2e^{i\rho_2},
\ena
where $E_3=|\epsilon_3|$ and $E_2=\pm|\epsilon_2|$. 
Then, the consistency conditions Eqs. (\ref{eq:ccnh2}) and (\ref{eq:ccih}) are expressed as
\bea
\left|
\frac{E_2 E_3}{k_3^2-|k_2|^2} 
~[ k_3\cos(\alpha_0 - \rho_2 - \rho_3) + |k_2|\cos(\rho_2 - \rho_3) ]~
\right| \simeq 0,
\label{eq:ccphco}
\ena
for the fully constructive interference, while
\bea
\left|
\frac{E_2 E_3}{k_3^2-|k_2|^2} 
~[ k_3\sin(\alpha_0 - \rho_2 - \rho_3) - |k_2|\sin(\rho_2 - \rho_3) ]~
\right| \simeq 0,
\label{eq:ccphde}
\ena
for the fully destructive interference.
Here, $k_3=1$ and $|k_2|=m_2^0/m_3^0=p_0$ for the NH case, while $k_3=m_3^0/m_2^0=1/p_0$ and $|k_2|=1$ for the IH case.
Suppose neither $E_3 = 0$ nor $E_2 = 0$, these conditions can be satisfied by adjusting the CP phases.
\begin{figure}[t]
\begin{center}
\includegraphics[width=7.0cm]{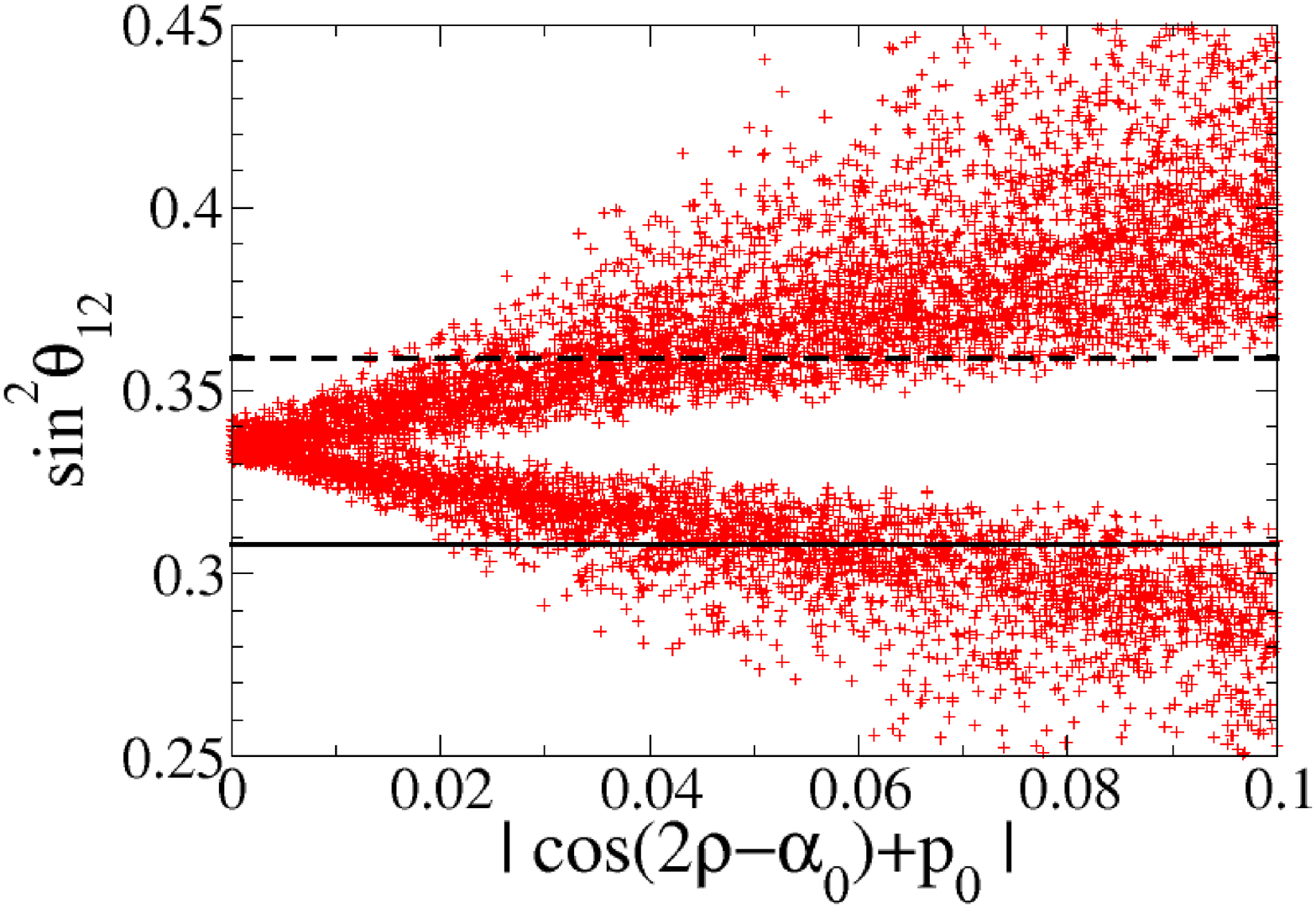}
\hspace{1.0cm}
\includegraphics[width=7.0cm]{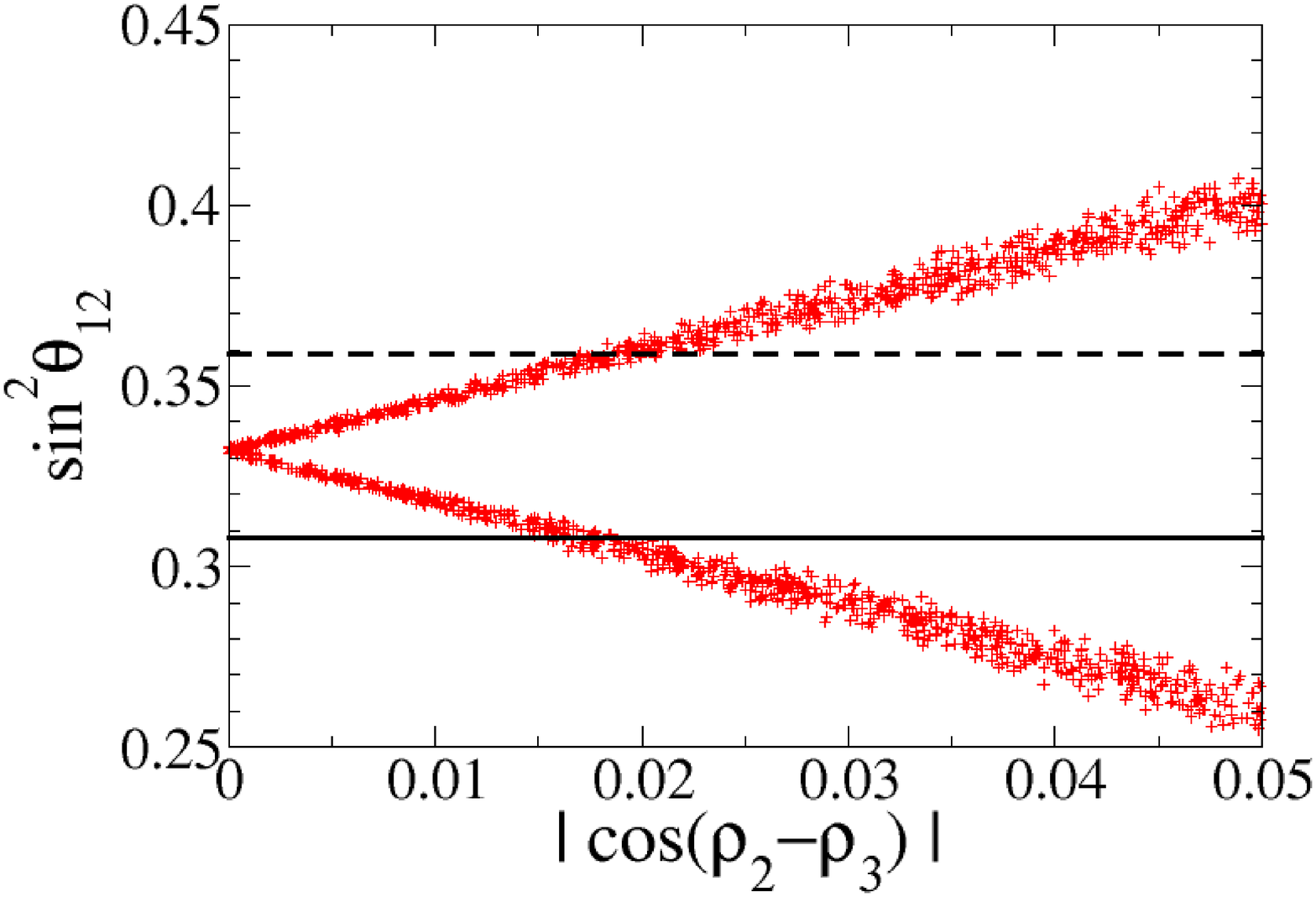}
\end{center}
\caption{\footnotesize 
Left panel: Scatter plot of $\sin^2\theta_{12}$ for the NH in the mass regions of $m_2 \simeq m_1$ ($p=0.4 - 0.8$), with $\cos\kappa = \pm 1$ and $\rho_2 = \rho_3$.
Right panel: Scatter plot of $\sin^2\theta_{12}$ for the IH case with $m_3 \simeq 0$ and $\cos\kappa = \pm 1$.
The horizontal dashed and solid lines display the $3\sigma$ upper bound and best-fit value, respectively.
}
\label{fig:ccph}
\end{figure}

\begin{itemize}
\item The NH case with the fully constructive interference and $\rho_2 = \rho_3 \equiv \rho$.\\
In this case, Eq. (\ref{eq:ccphco}) provides us with
\bea
|\cos(2\rho - \alpha_0)+p_0| \simeq 0.
\label{eq:ccnh2p}
\ena
In the left panel of Fig. \ref{fig:ccph}, we numerically diagonalize the mass matrix and plot $\sin^2\theta_{12}$ as a function $|\cos(\alpha_0-2\rho)+p_0|$ for $E_2 > 0$.
It can be seen that $\sin^2\theta_{12}$ approaches the TBM value as $|\cos(\alpha_0-2\rho)+p_0|$ gets close to zero.

By substituting $\cos(2\rho - \alpha_0)=-p_0$ into $f$ and $g$ in Eq. (\ref{eq:fg}), we find
\bea
f&=&\frac{E_3}{\sqrt{1-p_0^2}}e^{i(\alpha_0-\rho-\pi/2)},\nonumber\\
g&=&\frac{E_2}{\sqrt{1-p_0^2}}e^{i(\alpha_0-\rho-\pi/2)}.
\ena
In turn, from the first identity of Eq. (\ref{eq:tbm-angles}), it is found that $\rho$ is given by the Dirac and Majorana phases as
\bea
\rho=\alpha_0-\delta\mp\pi/2,
\ena
where $\pm$ stems from the sign of $V_{13}$, yielding
\bea
\cos(\alpha_0-2\delta) \simeq p_0.
\ena
Since $p_0\simeq p$ and $\alpha_0\simeq \alpha$, this is the relation among 
observables and the prediction of this model.
Furthermore, $\sin\theta_{13}$ and $\sin\theta_{23}$ are expressed as
\bea
|\sqrt{2} E_3 +E_2|&=&\sqrt{3(1-p_0^2)}\sin\theta_{13},\nonumber\\
|-E_3+\sqrt{2}E_2|&=&\sqrt{6(1-p_0^2)}\left| 
\frac{\sin\theta_{23}-1/\sqrt{2}}{\cos\delta} \right|.
\ena

Let us emphasize two more-simplified models.
(1) If both $\epsilon_3$ and $\epsilon_2$ are real, i.e., $\rho=0$, the Majorana CP phase $\alpha$ is directly related to the Dirac 
CP phase $\delta$ via
\bea
\alpha=\delta\pm\pi/2,
\ena
and also $\delta$ is related to $p$ as
\bea
p=\mp\sin\delta.
\ena
(2) If $\epsilon_3=\sqrt{2}\epsilon_2$ (thus, $E_3=\sqrt{2}E_2$) the model predicts the maximal $\theta_{23}$. 
This prediction is favored by the latest data of $\nu_\mu$ disappearance reported by the T2K experiment \cite{maxT2K}.

\item The IH case.\\
The same situation, $\rho_2 = \rho_3$ and $\beta_0 = \alpha_0$, cannot be applied for the IH case because it leads to
\bea
\left|
\frac1{p_0}\cos\left(2\rho-\alpha_0 \right) + 1
\right|  \simeq 0 ,
\ena
which cannot be satisfied because $(1/p_0)<1$.\footnote{It is possible to consider $\rho_2 = \rho_3$ for the fully destructive interference case.}
This seems to be quite a strong constraint for model-building.

Instead, it may be interesting to consider the case of $m_3 \simeq 0$, since a massless active neutrino can naturally be explained by considering two-right-handed-neutrino seesaw scenarios \cite{2RHN1,2RHN2,2RHN3,2RHN4}.
In this case, the consistency conditions become
\bea
| \cos(\rho_2-\rho_3) | \ll 1 
~~~~{\rm or}~~~~\rho_2 - \rho_3 \simeq \frac{\pi}{2}~~{\rm mod}~~\pi
\ena
for the fully constructive interference, while 
\bea
| \sin(\rho_2-\rho_3) | \ll 1 
~~~~{\rm or}~~~~\rho_2 - \rho_3 \simeq 0~~{\rm mod}~~\pi
\ena
for the fully destructive interference.
As can be seen, they suggest correlations between $\rho_3$ and $\rho_2$.
Conversely, it can be said that one can naturally satisfy the consistency conditions once the phase correlations are explained by model-building.
The scatter plot of $\sin^2\theta_{12}$ for the fully constructive interference case is displayed in the right panel of Fig. \ref{fig:ccph}.

\end{itemize}

\section{Concluding remarks}
We have seen that some correspondences exist between 
the constraint on input parameters and the output constraints on 
experimental observables. It is amazing that this kind of 
correlation is observed analytically as we have illustrated. 
The feature we saw here is a general one for models where 
we start from the neutrino mass matrix which reproduce experimental 
data well and reproduce the data by adding small perturbation terms 
in the presence of the degeneracy between $m_2$ and $m_1$. Our method 
will be useful for model-building.

\end{document}